\documentclass[sigconf]{acmart}

\usepackage{xspace}
\usepackage{listings}

\newcommand{\projectname}{\textit{Niffler}\xspace}
\AtBeginDocument{%
  \providecommand\BibTeX{{%
    \normalfont B\kern-0.5em{\scshape i\kern-0.25em b}\kern-0.8em\TeX}}}

\setcopyright{acmcopyright}
\copyrightyear{Preprint - 2020}
\acmYear{2020}

\acmDOI{XXXXX}

\acmConference[Preprint]{Preprint}{August 5, 2020}{Emory University, GA, USA}
\acmBooktitle{Preprint}

\acmPrice{}

\begin{document}

\title{A DICOM Framework for Machine Learning Pipelines against Real-Time Radiology Images}


%\author{Pradeeban Kathiravelu}
%\affiliation{%
%  \institution{Emory University}
%  \city{Atlanta}
%  \country{GA, USA}}
%\email{pradeeban.kathiravelu@emory.edu}

\author{Pradeeban Kathiravelu$^\dagger$, Puneet Sharma$^\dagger$, Ashish Sharma$^\dagger$, Imon Banerjee$^\dagger$, Hari Trivedi$^\dagger$, Saptarshi Purkayastha$^\diamond$, Priyanshu Sinha$^\ddagger$, Alexandre Cadrin-Chenevert$^\star$, Nabile Safdar$^\dagger$, Judy Wawira Gichoya$^\dagger$}
\affiliation{
\institution{$^\dagger$Emory University, Atlanta, GA, USA, 
$^\diamond$Indiana University Purdue University, Indianapolis, IN, USA,\\$^\ddagger$Mentor Graphics India Pvt. Ltd., Noida, India, $^\star$CISSS Lanaudiere, Laval University, Quebec City, Canada}
}
\renewcommand{\shortauthors}{Kathiravelu, et al.}

\begin{abstract}
Real-time execution of machine learning (ML) pipelines on radiology images is hard due to limited computing resources in clinical environments, whereas running them in research clusters requires efficient data transfer and processing capabilities. We propose \projectname, an integrated framework that enables the execution of ML pipelines at research clusters by efficiently querying and retrieving radiology images from the Picture Archiving and Communication Systems (PACS) of the hospitals. \projectname uses the Digital Imaging and Communications in Medicine (DICOM) protocol to fetch and store imaging data and provides metadata extraction capabilities and Application programming interfaces (APIs) to apply filters on the images. \projectname further enables the sharing of the outcomes from the ML pipelines in a de-identified manner. \projectname has been running stable for more than 19 months and has supported several research projects at the department. In this paper, we present its architecture and three of its use cases: an inferior vena cava (IVC) filter detection from the images in real-time, identification of scanner utilization, and scanner clock calibration. Evaluations on the \projectname prototype highlight its feasibility and efficiency in facilitating the ML pipelines on the images and metadata in real-time and retrospectively.
\vspace{-0.4em}

\end{abstract}

%\ccsdesc[300]{Applied computing~Bioinformatics}



\maketitle

%\vspace{-1em}

\section{Introduction}
\label{sec:intro}
The growing ML research in radiology highlights the potential for real-time processing of images from the hospital scanners. Radiology departments consist of several clinical systems such as PACS~\cite{huang1988picture} and Vendor-Neutral Archives (VNAs)~\cite{pantanowitz2018twenty} that receive images real-time from various scanners. A medical professional requires a certain amount of time to access and process the images produced by the scanners. Such diagnosis can be automated and accelerated, by extracting and processing the imaging data and their textual metadata from the healthcare images in the PACS with the help of ML pipelines. However, clinical systems have limited processing and memory resources to execute ML pipelines on radiology images efficiently~\cite{cho2015medical}. Biomedical informatics (BMI) research clusters and cloud environments are designed for heavy computing workload. Despite the advancements in ML frameworks on medicine, a real-time big data processing framework, spanning a hospital network to a research cluster, is still lacking.

Patient wait times for examination and diagnosis significantly impacts the effectiveness of patient care~\cite{soremekun2011framework}. As radiologists are often a scarce resource, their availability affects the wait times. Rapid progress in the last decade in computer vision and natural language processing has ignited imaginations that Artificial Intelligence (AI) will lead to lower costs, fewer errors, more efficiency, and better health care~\cite{noorbakhsh2019artificial}. ML pipelines have been proposed to reliably prognosis and predict cancer to mitigate the workload of the diagnostic radiologists. Furthermore, scanner optimization and effective scheduling of patients to the available scanners can significantly reduce the wait times. Real-time processing of images and their metadata can facilitate such scheduling optimizations, by running computations and analytics on system performance metrics.

Several factors should be satisfied to run the computational pipelines in real-time on research clusters. First, there should a fast and secure data transfer from the PACS to the research clusters. Second, an efficient processing framework must be built to process the received images, facilitating the execution of ML pipelines on them. Enabling secured transfer and access to large volumes of data to be processed by the algorithms is mandatory for such executions. DICOM standardizes the format the healthcare imaging and communications are stored and transferred across the network~\cite{parisot1995dicom, pianykh2009digital}. DICOM network protocol facilitates the reliable transfer of imaging data and Structured Reports (SR)~\cite{clunie2000dicom} between the PACS and computing servers such as data centers, clouds, and research clusters. However, running the workflows on third-party public clouds comes with privacy concerns on sensitive healthcare data. Therefore, executing the ML pipelines on research clusters is often the only viable and secure option for healthcare images with protected health information (PHI).

This paper presents \projectname, an ML framework that retrieves images from the PACS using DICOM network listeners, and extracts and processes metadata from the acquired images at the research clusters. It then executes ML pipelines and real-time analytics pipelines on radiology images and their textual metadata. The data retrieval includes a push-based real-time data transfer and query-driven specific data pulls, to analyze both real-time and retrospective studies. We have demonstrated the capability and stability of \projectname in receiving data in real-time securely, by running it continuously over 19 months on the research clusters to receive images from two PACS. The \projectname prototype deployment receives real-time data from one PACS while retrieving query-based data from an enterprise archive PACS that stores historical data. Thus, \projectname facilitates inference pipelines for the clinical validation of AI models, enabling Real-Time Analytics (RTA)~\cite{trinks2017real}.

We propose and prototype three use cases of \projectname. The first use case is an IVC filter detection on radiology images of modality XR, DX, CR, DR, and DX CR, for chest, spine, abdomen, small intestines, gallbladder, and thorax. We used the RetinaNet object detection~\cite{lin2017focal} as our IVC filter detection model on the images. The ML pipeline understands the context of care for patients with IVC filters, including whether the patients are anticoagulated, their anticoagulation profile, and if they have an upcoming filter retrieval appointment. We observe high accuracy and efficiency with \projectname in IVC filter detection. The second use case is computing scanner utilization by performing computations on metadata. We were able to calculate the scanner utilization more accurately with \projectname, compared to the clinical data warehouse (CDW)~\cite{grant2006integrating}. The third use case finds the scanner clock miscalibrations by comparing the time the real-time images are received in the research cluster with the time the images acquired by the scanner, as indicated in the metadata.

%\vspace{-1em}
\section{Background and Motivation}
\label{sec:background}

Open-source frameworks such as PyTorch~\cite{ketkar2017introduction} and Tensorflow~\cite{abadi2016tensorflow}, as well as the pre-trained models released through various model zoos, have catalyzed the maturation frameworks for Artificial Intelligence (AI) training. Multiple systems are now available for testing AI models in real-life settings. A review of the exhibits at the annual radiology meeting – RSNA 2019 shows these tools are developed as market places for already approved FDA algorithms or systems embedded into existing PACS and imaging equipment like chest Xrays. The American College of Radiology has developed an open-source system called the ACR AI-LAB~\cite{acr} to democratize AI among radiologists by supporting on-premise federated learning. \projectname aims to execute ML pipelines in the CPUs and GPUs readily available in the research clusters on the radiology images retrieved in real-time. It facilitates the validation of the ML pipelines and integration with systems such as the ACR AI-LAB.

Real-time decision-making pipelines for radiologists support triage of emergent studies and reduce the rate of addendums~\cite{burns2020just}. It also supports direct calls from the referring doctor to the correct person, such as calls to technologists when a study is ordered and not performed, and to the reading room when a study has been dictated. Researchers have built a web-based framework for quantitative imaging informatics~\cite{epad}. The framework offers radiology image metadata with the Annotation and Image Markup (AIM) standard~\cite{mongkolwat2014national}, and implements AIM as a web service with semantic image annotation. It is extensible but is limited in its capabilities. For example, a de-identification pipeline cannot be attached to it. Rather, DICOM images must be manually de-identified to provide anonymization before loading them to the web service. Through its extensible and compact architecture, \projectname also supports the integration of features such as de-identification and conversion between various image formats.

Despite advances in AI in medicine, there is a low rate of validation of medical imaging AI studies where only 6\% (31/516) of published studies in 2018 performed external validation (i.e., diagnostic cohort, multi-institution data, prospective)~\cite{kim2019design}. Even the FDA's streamlined AI approval process often only requires a few test samples, with further testing viewed as post-marketing surveillance~\cite{fda1,fda2}. Beyond infrastructure for training algorithms, there is still a critical gap in providing tooling to help inference evaluation of AI systems in local systems before large scale deployment. Real-life testing of ML pipelines can take months to set up after the initial training, with time invested in developing imaging streams that are not in production, security requirements for on-premise deployment and with minimal disruption on the workflow. Among 20 ML articles in Nature Medicine in 2018/2019, only one had a graphical user interface for model interaction, while ten provided code that requires significant time to set up~\cite{abid2020online}. \projectname aims to provide a framework that is easy to configure and use, to retrieve images and run ML pipelines natively on the research clusters with its support for containerized ML workflows.

AI models are brittle, and they do not generalize. Dataset shift refers to original training data characteristics change, causing declines in AI performance over time~\cite{subbaswamy2020development}. It requires continuous monitoring and recalibration. Differences in radiology equipment within/across institutions affect generalizability, and a model can learn and fine-tune itself based on equipment-specific details, affecting performance and clinical utility~\cite{zech2018variable}. Clinical features like chest tubes for pneumothorax undermine model performance, as they detect the tubes rather than pneumothorax~\cite{taylor2018automated}. \projectname queries and retrieves DICOM data with minimal latency, and facilitates subsequent extraction and processing of imaging data and the textual metadata. Thus, \projectname supports continuous monitoring and recalibration for various ML workflows for multiple studies based on pre-defined filters.

There is a gap between engineering metrics to evaluate algorithms, and what is clinically useful. Model results are typically presented as confusion matrices or ROC (Receiver Operating Characteristic) curves, but they do not translate to clinical use. Identifying sampling biases usually requires manual review and domain expertise and may not be apparent during model testing before actual images are reviewed. Most significantly, radiologists must participate in AI development, collect test cases, establish ground truth, choose appropriate metrics and performance thresholds, and evaluate test cases with continuously monitoring outputs. Our vision for \projectname is to develop and test an AI inference pipeline that combines clinical and imaging data. Thus, we aim to facilitate radiologists' participation in AI clinical validation.

%\vspace{-1.5em}
\section{The \projectname Framework}
\label{sec:arch}

We designed \projectname as a framework that retrieves DICOM images real-time and on-demand from PACS to a research cluster. By extracting and analyzing the metadata at the research clusters, \projectname enables the creation of image subsets that can be further processed, used as data for ML workflows, or shared with other researchers. 

\subsection{Architecture}

\begin{figure}[!ht]
%\vspace{-1em}
    \begin{center}
  \centering 

        \resizebox{\columnwidth}{!}{
            \includegraphics[width=\textwidth]{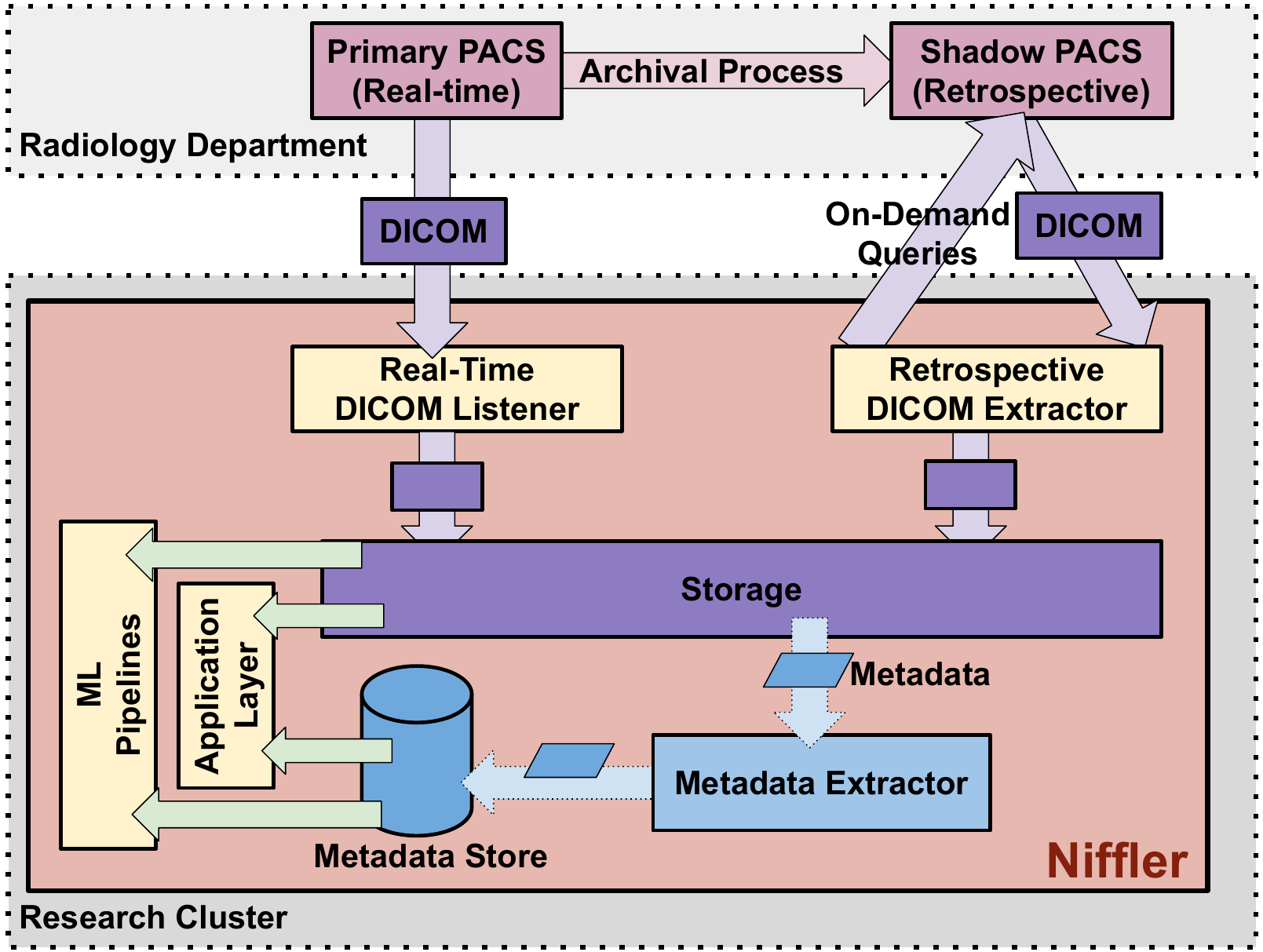}
        }
    \end{center}
%  \vspace{-1em}

  \caption{Deployment Architecture}
%  \vspace{-1em}
  \label{fig:rpacs} 
\end{figure}

Figure~\ref{fig:rpacs} depicts a sample \projectname deployment. \projectname consists of instances of DICOM listeners for receiving DICOM images real-time, and retrospective DICOM extractors to query and retrieve images on-demand. It consists of a Metadata Extractor that extracts the textual metadata from the retrieved DICOM images. \projectname stores the images in its storage and the metadata in a Metadata Store. Its Application Layer provides unified access to data and metadata in the storage and the metadata store. Thus, ML and processing pipelines run efficiently on the images and metadata stored in \projectname.

In the standard healthcare system, the radiology department may consist of several PACS, each receiving radiology images from scanners of various modalities. Our current deployment environment consists of 2 PACS from our institutional radiology department, configured to accept DICOM retrieval queries from \projectname. We deployed \projectname in a standard server with 12 GB memory and 1 TB  of hard disk, in the research cluster. In this sample deployment, the primary PACS receives data in real-time from the scanners. The radiology department has configured an archival process that copies the images from the primary PACS to the shadow PACS and then cleans up the primary PACS every week. Hence, the shadow PACS stores imaging data for several years, supporting retrospective queries.

The real-time DICOM listener receives images from the primary PACS continuously as a DICOM imaging stream. The retrospective DICOM extractor performs on-demand queries issued by the users on the shadow PACS. \projectname consists of multiple DICOM StoreSCP processes configured at the research cluster, one for each PACS. The images from the PACS are stored separately in a storage, in a hierarchical folder structure: patient-folder/study-folder/series-folder/instance.dcm.

The Metadata Extractor executes its extraction query on all the images in the storage, extracts the relevant metadata from the DICOM headers, and stores the metadata free from PHI in a Metadata Store. The Application Layer facilitates access to the DICOM images from the storage and the respective metadata from the metadata store. It provides utility functions such as de-identification, image conversion, and scripts such as scanner utilization computation and scanner time calibration. The ML pipelines run either directly or via the application layer on the images and the metadata. \projectname deletes the images from the storage periodically once the metadata extraction and the execution of the ML pipelines on the images are complete. Subsets of images relevant for a study can be shared with the other researchers, typically after processing them such as, de-identifying them, converting them into png images, or after running the ML pipelines on them together with their results.

The metadata store uses a NoSQL~\cite{han2011survey} database due to its scalability and support for data in JSON format for the hierarchical format of DICOM metadata. The database consists of several collections, each representing a \textit{profile}. Each profile represents the DICOM attributes that must be extracted for one or more experiments. The Metadata Extractor reads the folder consisting of the \textit{profiles} stored as text files. It parses the DICOM headers from the images received in the storage, and stores the relevant attributes defined in the profiles into their respective collections. An experiment can use existing profiles or create a new profile at run time without halting the execution. As each profile creates a collection, the access can be limited to the respective researcher at the collection level, using the access controls offered by the NoSQL metadata store. The metadata is used to filter cohorts and sub cohorts that allow dataset creation for model inference. For example, to test whether an IVC filter model performance drops with the change of equipment, cohorts of data filtered by modality, and manufacturer are easily created at the metadata level.

The application layer provides a unified data explorer access to both data and metadata. It allows the users to determine the cohort components required for model inference. For example, an end-user will access the metadata (without specific clinical information) and filter with a query like ``I want all Abdomen X-rays for studies between 2012 and 2019 with sub cohorts of manufacturers and their anticoagulation medication and problem list''. Since the pipeline is a prospectively populated system with an option for a query to extract images meeting a specific criterion, this limits the amount of information stored in the research clusters that are duplicated. Without the integrated pipeline, a researcher would have to submit multiple queries to the PACS research team and CDW, work on anonymizing the data collected, merge the data, and then run the model inference. Thus, \projectname supports prospective dynamic cohort and subcohort creation, eliminating the need for duplicate data storage and aggregation, with anonymized model output. Through its cloud-native architecture that natively supports the execution of pipelines as containers, \projectname provides an infrastructure-agnostic execution with seamless scaling and migration.

%\vspace{-1em}

\subsection{\projectname Execution}
\projectname autostarts at login as a service with its storage and metadata extraction processes. \projectname Metadata Extractor opens a \textit{Pickle} file that consists of a set of DICOM series: $P$ -- set of series whose metadata is already extracted but the images are not deleted, and $D$ -- set of series that are already processed and deleted. If the Pickle file does not exist, the sets are initialized as empty sets, and the Pickle file is created.

At the core of the Metadata Extractor is an \textit{extract\_metadata} process that runs periodically (by default, set at every 10 minutes) as a thread. However, only one instance of \textit{extract\_metadata} is run at any given time. It traverses all the DICOM series ($S$) in the storage. As the data is, by default, stored in the file system, the Metadata Extractor uses the $find$ operating system command, as a sub-process. In each iteration, the Metadata Extractor extracts metadata from the first image of each series that is not extracted yet. For performance reasons, \projectname extracts metadata from only one image per series. However, we can configure it to extract more than one (such as first, last, and a middle instance in any given series) or all the images of each series. The sets are updated as depicted in (\ref{eq:co1n}) following an iteration of extraction.

\begin{equation} \label{eq:co1n}
\forall s \in S, s \notin P \wedge s \notin D \implies P \gets P \cup \{s\}
\end{equation}

\projectname has a periodic \textit{clear\_storage} process. By default, this process runs once at 23:59 each night. It deletes the images whose metadata is already extracted, making sure no ML or processing workflow is processing them. The sets are updated as depicted in (\ref{eq:co1d}) following an iteration of deletion.

\begin{equation} \label{eq:co1d}
\forall s \in P \implies D \gets D \cup \{s\}, P \gets P \setminus \{s\}
\end{equation}

An \textit{update\_pickle} process runs periodically (by default, every 20 minutes), to ensure that the sets $P$ and $D$ are written to the Pickle file. This approach of writing the sets to the filesystem and reading them upon startup ensures that the Metadata Extractor processes resume where they stopped when \projectname halts involuntarily or stopped for updates. This state-awareness aims to support seamless updates and improve fault-tolerance, ensuring that the progress made by the extraction and deletion processes is not lost upon failures and restarts.

\subsection{Implementation}

We developed the \projectname prototype as an open-source platform\footnote{The source code can be found at \url{https://github.com/Emory-HITI/Niffler}}, with its core in Python3. \projectname uses the Pydicom library to extract metadata and process the DICOM images. The application layer and the ML pipelines are developed in multiple languages. The application layer consists of several toolkits. Among these, the scanner utilization is in Java, whereas scanner timeshift calibration is in Javascript. Among the ML pipelines, the IVC filter detection container is developed in Python3.

Two instances of the DCM4CHE~\cite{warnock2007benefits} StoreSCP tool is configured to receive all images in 2 different ports (4242 and 4243). The first one listens to all the images sent from the real-time PACS and accepts them. The second one receives the images on-demand via CM4CHE MoveSCU queries. The DCM4CHE MoveSCU is configured to retrieve images based on specific queries, which are optionally first filtered based on a FindSCU. The \projectname prototype uses the local file system as the storage and MongoDB~\cite{banker2011mongodb} as its Metadata Store. We configured a replicated MongoDB cluster to support the scaling and redundancy of the metadata store, as mongo replicasets can be added to the MongoDB cluster without reconfiguring the database. The folders that store the DICOM images are identified as their unique IDs such as PatientID, StudyInstanceUID, and SeriesInstanceUID, and thus indexed and easily identifiable from the metadata. Given below is a sample (anonymized) entry in the metadata store for a DICOM image.

{\fontsize{8}{8}\selectfont                  
\begin{lstlisting}
{"_id" : ObjectId("5e7e1d1a58a2a12adff64811"),
"PatientID" : "XXXX",
"StudyInstanceUID" : "XXXX",
"StudyDate" : "20200327",
"StudyTime" : "050348.128",
"SeriesInstanceUID" : "XXXX",
"SeriesDate" : "20200327",
"SeriesTime" : "050503.271",
"SOPInstanceUID" : "XXXX",
"AcquisitionDate" : "20200327",
"AcquisitionTime" : "050503.271",
"Exposure" : "1",
"ExposureTime" : "7",
"ImageType" : ["DERIVED",
"PRIMARY"
],
"Modality" : "DX",
"StationName" : "XXXX",
"StudyDescription" : "XR CHEST 1 VIEW PORTABLE",
"InstitutionName" : "XXXX",
"SeriesNumber" : "2",
"SeriesDescription" : "AP",
"BodyPartExamined" : "PORT CHEST",
"DeviceSerialNumber" : "002691"
}
\end{lstlisting}
}
%\vspace{-0.5em}

We deployed \projectname in a server secured by strict firewall rules and configured the MongoDB instances with authentication. For data transfer efficiency, \projectname supports receiving data in a secure compressed DICOM data stream. In our sample deployment, the images received from the PACS are in JPEG lossless compressed form. \projectname uses GDCM~\cite{developers2010grass} to export the compressed DICOM images to a PNG format, for the ML pipelines to consume in a de-identified manner. \projectname supports running ML pipelines as Docker~\cite{merkel2014docker} containers on the images and metadata that it stores. By supporting ML pipelines as Docker containers, \projectname minimizes the repetitive and complicated configuration steps while automating the end-to-end process with seamless deployment.
\section{Evaluation}
\label{sec:eval}

\projectname extracted data from 715 scanners, receiving DICOM data up to 350 GB each day and continuously running for more than 19 months. \projectname has facilitated several ML workflows in our department. In this section, we will look into three such use cases to evaluate the performance of \projectname supporting radiologists ML and processing workflows in real-time.

\subsection{IVC Filter (IVCF) Detection}

\begin{figure}[!ht]
    \begin{center}
    %\vspace{-1em}
  \centering 

        \resizebox{\columnwidth}{!}{
            \includegraphics[width=\textwidth]{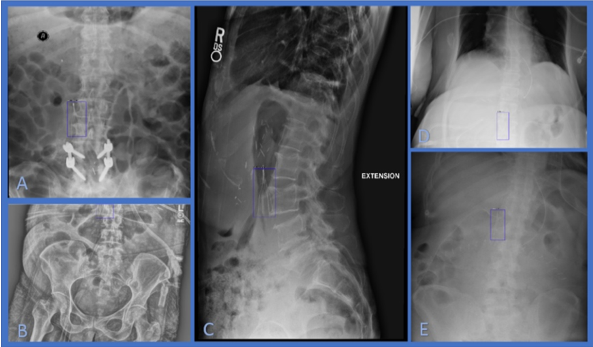}
        }
    \end{center}
    %\vspace{-1em}

  \caption{IVCF Detection and Localization on Various Views}
  \label{fig:ivc} 
\end{figure}

To measure the performance efficiency and viability of \projectname, we built an IVCF detection pipeline as a container to execute on the DICOM retrieved in real-time with \projectname. The pipeline uses the Keras RetinaNet object detection pre-trained model to determine whether an IVCF is detected in the subcategories of the images. The backbone encoder CNN was based on the Resnet-50 architecture~\cite{he2016deep} pre-trained on the COCO object detection dataset~\cite{lin2014microsoft}. The model was trained on 503 abdominal, thoracoabdominal, and lumbar radiographs from various projection views and validated on 127 images. During the real-time inference, first the \projectname Metadata Extractor applied the filters on modality and body parts to create a DICOM subset, consisting of 989 DICOM images. The IVCF detection container ran its inference on the identified images, including chest Xray, abdomen radiographs, and Spine Xrays. The pipeline drew a bounding box around the identified IVCF in the images and outputs a PNG image with the detection box, as Figure~\ref{fig:ivc} shows.Two interventional radiologists reviewed all the outputs and determined that the IVCF detection algorithm classified the test images with high accuracy of 96.0\% on the 989 test images retrieved in real-time with \projectname.

As we receive the images in real-time, the Metadata Extractor applies the filters on modality and body parts to create a subset of data. The object detection algorithm of the ML pipeline executes, taking the identified images as its input. The IVCF detection container first converts the DICOM images into PNG images before running its inference on the PNG images, including chest Xray, abdomen radiographs, and Spine Xrays. Then the pipeline draws a bounding box around the filter and outputs a PNG image with the detection box, as shown in Figure~\ref{fig:ivc}. The Retinanet IVCF detection algorithm classified the test images with high accuracy of 96.0\% on the images retrieved in real-time with \projectname. 

%\vspace{-2em}

\subsection{Calculating Scanner Utilization}
\begin{figure}[!ht]
%\vspace{-1em}
    \begin{center}
  \centering 

        \resizebox{\columnwidth}{!}{
            \includegraphics[width=\textwidth]{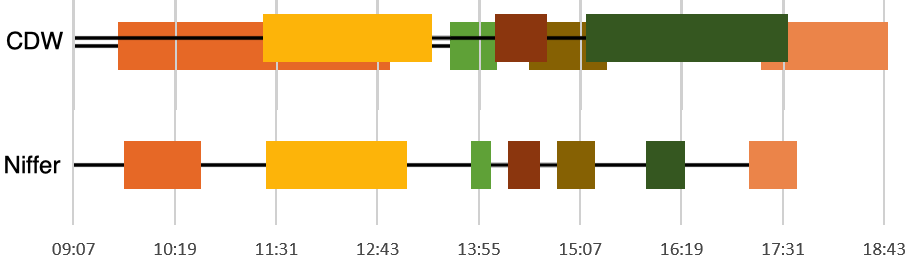}
        }
    \end{center}
%\vspace{-1em}

  \caption{Visualizing Scanner Utilization Measurements}
  \label{fig:su} 
  %\vspace{-1em}

\end{figure}

Next, we demonstrate the potential to use DICOM metadata to understand the scanners' system performance metrics for operational efficiency, individually for each scanner, and for the healthcare networks to schedule the patients efficiently across the scanners and hospitals. \projectname calculates the time each scanner is used for an examination, using the AcquisitionTime attribute in the DICOM metadata. Then it computes the scanner utilization at any time of the day by dividing the calculated examination time by the total time the scanner is turned on for the day. We calculated scanner utilization by using the metadata from the DICOM images received real-time by \projectname for a day, for a specific scanner. We then compared it against the data derived from CDW. Figure~\ref{fig:su} indicates the time windows a scanner is used for examinations, as indicated by \projectname and CDW.

We observe that \projectname more accurately identified when a scanner is used in an examination and when it is idling in between the examinations. CDW is more prone to human errors. It reported overlapping examinations for a scanner, which is not possible as a scanner can perform no more than one examination at any time. \projectname correctly identified the scanner utilizations as the metadata consists of the actual acquisition time. Consequently, using \projectname, we can evaluate the scanner performance and optimize how the scanners are utilized. For instance, combined with the clinical data, we may identify why certain scanners have longer wait times between examinations and whether a longer examination time is justified or necessary to provide the acquired image with acceptable quality.

%\vspace{-1em}
\subsection{Scanner Clock Calibration}
Finally, we used \projectname to identify the scanners that have misconfigured time, using the metadata of the images received in real-time. When metadata attributes such as AcquisitionTime and SeriesTime have a wrong timestamp due to the scanner time misconfiguration (typically, due to a wrong timezone setting), extracting useful information from the metadata becomes harder. \projectname contains a script that compares the time the images are received at the metadata store against the acquisition time from the metadata. Typically the difference does not exceed 20 minutes, accounting for the time the image to be received over the network and the metadata to be extracted and saved. \projectname correctly identified five scanners with misconfigured time.

Our evaluations highlight that \projectname provides fast processing of ML models in real-time with high-efficiency, by running the filtering at metadata level and the ML pipelines using CPU only on the identified images. We also highlight the potential in using \projectname in operating the scanners better, with information not otherwise readily available in the clinical systems.

\subsection{Discussion}
The prototype and the evaluations highlight the potential and performance of \projectname in executing ML pipelines and metadata processing workflows in real-time in a research cluster. However, \projectname requires extension and further development for clinical validation.

In the IVCF detection, we currently do not know if the patient is anticoagulated and can have this filter removed, or whether they have contraindication of filter removal or already have an upcoming scheduled appointment for filter retrieval. As future work, we propose to support end-to-end clinical validation of the ML pipelines with the consumption of Electronic Medical Record (EMR) from the RTA on the laboratory information (INR, anticoagulation profile), medications (whether the patient is on any anticoagulant), problem list (for example, if a patient has a history of GI bleed and hence cannot be anticoagulated) and the upcoming clinical appointments where a patient can be seen in the clinic. Linkage to an ADT message would allow just in time clinical review of the patients in same-day appointments. In the IVCF detection pipeline, such a linkage will provide education to providers on the benefits of the IVCF removal when no longer required.

By merging the \projectname real-time images with the RTA stream, we can create a live AI inference pipeline that accelerates the development of clinically useful algorithms. To our knowledge, this will be the first AI inference pipeline that combines real-time image and clinical data information during AI validation. We propose integrating the RTA clinical data pipeline with the imaging pipeline to provide tooling for data curation for model inference and training. Specifically, \projectname will receive an HL7 feed from the RTA system, which will be normalized into Fast Healthcare Interoperability Resources (FHIR)~\cite{bender2013hl7} resource groups. For the IVCF detection, we will limit the scope of integration to cover the following resource groups – Patient, Organization, Appointment and Schedule, Medication, and Observation. The FHIR resource groups will be added to our existing DICOM metadata and stored as MongoDB collections in the metadata store. 
%\vspace{-0.5em}
\section{Conclusion}
\label{sec:concl}
In this paper, we presented \projectname, a framework that supports the seamless transfer of data from the PACS to the research clusters and enables efficient execution of ML pipelines on the images, reports, and the extracted textual metadata. \projectname facilitates the execution of ML models with a minimal tuning of infrastructure. It further enables the development of models against real-time data streams and helps gather large-scale prospective data in a centralized store to facilitate imaging research. We demonstrated the potential for seamless execution of ML pipelines in real-time with three use cases of \projectname -- one on ML workflows on the DICOM images, and the other two by processing the extracted metadata. Our evaluations and prolonged execution of the \projectname prototype highlight its support for efficient processing of data.

As future work, we propose to add de-identification pipelines from The Cancer Imaging Archive (TCIA)~\cite{clark2013cancer}, allowing human curation and a centralized process for IRB, thus ensuring that the data pipelines produce high-quality data in a secure and scalable manner. We propose using standardized instruments of the unified theory of acceptance and use of technology (UTAUT)~\cite{venkatesh2016unified} to measure the perceived usefulness and satisfaction of the inference pipeline for AI validation. We will use the applied cognitive task analysis (ACTA)~\cite{militello1998applied} previously validated for extracting task information from subject matter experts and adapt it for clinical validation. This approach will entail a task interview with think aloud observations as radiologists validate AI models on the inference pipeline, followed by a knowledge audit of the decision making for determining clinical performance and utility of the ML pipelines. 

%\vspace{-0.5em}

\bibliographystyle{ACM-Reference-Format}
\bibliography{sample-base}


\begin{thebibliography}{35}

%%% the Web address.

\ifx \showCODEN    \undefined \def \showCODEN     #1{\unskip}     \fi
\ifx \showDOI      \undefined \def \showDOI       #1{#1}\fi
\ifx \showISBNx    \undefined \def \showISBNx     #1{\unskip}     \fi
\ifx \showISBNxiii \undefined \def \showISBNxiii  #1{\unskip}     \fi
\ifx \showISSN     \undefined \def \showISSN      #1{\unskip}     \fi
\ifx \showLCCN     \undefined \def \showLCCN      #1{\unskip}     \fi
\ifx \shownote     \undefined \def \shownote      #1{#1}          \fi
\ifx \showarticletitle \undefined \def \showarticletitle #1{#1}   \fi
\ifx \showURL      \undefined \def \showURL       {\relax}        \fi
% invisible to TeX
\providecommand\bibfield[2]{#2}
\providecommand\bibinfo[2]{#2}
\providecommand\natexlab[1]{#1}
\providecommand\showeprint[2][]{arXiv:#2}

\bibitem[\protect\citeauthoryear{Abadi, Barham, Chen, Chen, Davis, Dean, Devin,
  Ghemawat, Irving, Isard, et~al\mbox{.}}{Abadi et~al\mbox{.}}{2016}]%
        {abadi2016tensorflow}
\bibfield{author}{\bibinfo{person}{Mart{\'\i}n Abadi}, \bibinfo{person}{Paul
  Barham}, \bibinfo{person}{Jianmin Chen}, \bibinfo{person}{Zhifeng Chen},
  \bibinfo{person}{Andy Davis}, \bibinfo{person}{Jeffrey Dean},
  \bibinfo{person}{Matthieu Devin}, \bibinfo{person}{Sanjay Ghemawat},
  \bibinfo{person}{Geoffrey Irving}, \bibinfo{person}{Michael Isard},
  {et~al\mbox{.}}} \bibinfo{year}{2016}\natexlab{}.
\newblock \showarticletitle{{Tensorflow: A system for large-scale machine
  learning}}. In \bibinfo{booktitle}{\emph{12th $\{$USENIX$\}$ Symposium on
  Operating Systems Design and Implementation ($\{$OSDI$\}$ 16)}}.
  \bibinfo{pages}{265--283}.
\newblock


\bibitem[\protect\citeauthoryear{Abid, Abdalla, Abid, Khan, Alfozan, and
  Zou}{Abid et~al\mbox{.}}{2020}]%
        {abid2020online}
\bibfield{author}{\bibinfo{person}{Abubakar Abid}, \bibinfo{person}{Ali
  Abdalla}, \bibinfo{person}{Ali Abid}, \bibinfo{person}{Dawood Khan},
  \bibinfo{person}{Abdulrahman Alfozan}, {and} \bibinfo{person}{James Zou}.}
  \bibinfo{year}{2020}\natexlab{}.
\newblock \showarticletitle{An online platform for interactive feedback in
  biomedical machine learning}.
\newblock \bibinfo{journal}{\emph{Nature Machine Intelligence}}
  \bibinfo{volume}{2}, \bibinfo{number}{2} (\bibinfo{year}{2020}),
  \bibinfo{pages}{86--88}.
\newblock


\bibitem[\protect\citeauthoryear{{ACR}}{{ACR}}{2020}]%
        {acr}
\bibfield{author}{\bibinfo{person}{{ACR}}.} \bibinfo{year}{2020}\natexlab{}.
\newblock \bibinfo{title}{{ACR AI-LAB}}.
\newblock
\newblock
\newblock
\shownote{Available at https://www.acrdsi.org/Get-Involved/AI-LAB.}


\bibitem[\protect\citeauthoryear{Banker}{Banker}{2011}]%
        {banker2011mongodb}
\bibfield{author}{\bibinfo{person}{Kyle Banker}.}
  \bibinfo{year}{2011}\natexlab{}.
\newblock \bibinfo{booktitle}{\emph{{MongoDB in action}}}.
\newblock \bibinfo{publisher}{Manning Publications Co.}
\newblock


\bibitem[\protect\citeauthoryear{Bender and Sartipi}{Bender and
  Sartipi}{2013}]%
        {bender2013hl7}
\bibfield{author}{\bibinfo{person}{Duane Bender} {and} \bibinfo{person}{Kamran
  Sartipi}.} \bibinfo{year}{2013}\natexlab{}.
\newblock \showarticletitle{HL7 FHIR: An Agile and RESTful approach to
  healthcare information exchange}. In \bibinfo{booktitle}{\emph{Proceedings of
  the 26th IEEE international symposium on computer-based medical systems}}.
  IEEE, \bibinfo{pages}{326--331}.
\newblock


\bibitem[\protect\citeauthoryear{Burns, Hasting, Gichoya, McKibben, Shea, and
  Frank}{Burns et~al\mbox{.}}{2020}]%
        {burns2020just}
\bibfield{author}{\bibinfo{person}{John~L Burns}, \bibinfo{person}{Dan
  Hasting}, \bibinfo{person}{Judy~W Gichoya}, \bibinfo{person}{Ben McKibben},
  \bibinfo{person}{Lindsey Shea}, {and} \bibinfo{person}{Mark Frank}.}
  \bibinfo{year}{2020}\natexlab{}.
\newblock \showarticletitle{{Just in Time Radiology Decision Support Using
  Real-time Data Feeds}}.
\newblock \bibinfo{journal}{\emph{Journal of Digital Imaging}}
  \bibinfo{volume}{33}, \bibinfo{number}{1} (\bibinfo{year}{2020}),
  \bibinfo{pages}{137--142}.
\newblock


\bibitem[\protect\citeauthoryear{Cho, Lee, Shin, Choy, and Do}{Cho
  et~al\mbox{.}}{2015}]%
        {cho2015medical}
\bibfield{author}{\bibinfo{person}{Junghwan Cho}, \bibinfo{person}{Kyewook
  Lee}, \bibinfo{person}{Ellie Shin}, \bibinfo{person}{Garry Choy}, {and}
  \bibinfo{person}{Synho Do}.} \bibinfo{year}{2015}\natexlab{}.
\newblock \showarticletitle{{Medical image deep learning with hospital PACS
  dataset}}.
\newblock \bibinfo{journal}{\emph{arXiv preprint arXiv:1511.06348}}
  (\bibinfo{year}{2015}).
\newblock


\bibitem[\protect\citeauthoryear{Clark, Vendt, Smith, Freymann, Kirby, Koppel,
  Moore, Phillips, Maffitt, Pringle, et~al\mbox{.}}{Clark
  et~al\mbox{.}}{2013}]%
        {clark2013cancer}
\bibfield{author}{\bibinfo{person}{Kenneth Clark}, \bibinfo{person}{Bruce
  Vendt}, \bibinfo{person}{Kirk Smith}, \bibinfo{person}{John Freymann},
  \bibinfo{person}{Justin Kirby}, \bibinfo{person}{Paul Koppel},
  \bibinfo{person}{Stephen Moore}, \bibinfo{person}{Stanley Phillips},
  \bibinfo{person}{David Maffitt}, \bibinfo{person}{Michael Pringle},
  {et~al\mbox{.}}} \bibinfo{year}{2013}\natexlab{}.
\newblock \showarticletitle{{The Cancer Imaging Archive (TCIA): maintaining and
  operating a public information repository}}.
\newblock \bibinfo{journal}{\emph{Journal of digital imaging}}
  \bibinfo{volume}{26}, \bibinfo{number}{6} (\bibinfo{year}{2013}),
  \bibinfo{pages}{1045--1057}.
\newblock


\bibitem[\protect\citeauthoryear{Clunie}{Clunie}{2000}]%
        {clunie2000dicom}
\bibfield{author}{\bibinfo{person}{David~A Clunie}.}
  \bibinfo{year}{2000}\natexlab{}.
\newblock \bibinfo{booktitle}{\emph{{DICOM structured reporting}}}.
\newblock \bibinfo{publisher}{PixelMed publishing}.
\newblock


\bibitem[\protect\citeauthoryear{Developers}{Developers}{2010}]%
        {developers2010grass}
\bibfield{author}{\bibinfo{person}{GDCM Developers}.}
  \bibinfo{year}{2010}\natexlab{}.
\newblock \bibinfo{title}{{Grass Roots DiCoM}}.
\newblock
\newblock


\bibitem[\protect\citeauthoryear{{ePAD}}{{ePAD}}{2020}]%
        {epad}
\bibfield{author}{\bibinfo{person}{{ePAD}}.} \bibinfo{year}{2020}\natexlab{}.
\newblock \bibinfo{title}{{ePAD: web-based platform for quantitative imaging in
  the clinical workflow }}.
\newblock
\newblock
\newblock
\shownote{Available at https://epad.stanford.edu.}


\bibitem[\protect\citeauthoryear{FDA}{FDA}{2020a}]%
        {fda2}
\bibfield{author}{\bibinfo{person}{FDA}.} \bibinfo{year}{2020}\natexlab{a}.
\newblock \bibinfo{title}{{Digital Health Software Precertification (Pre-Cert)
  Program | FDA}}.
\newblock
\newblock
\newblock
\shownote{Available at
  https://www.fda.gov/medical-devices/digital-health/digital-health-software-precertification-pre-cert-program.}


\bibitem[\protect\citeauthoryear{FDA}{FDA}{2020b}]%
        {fda1}
\bibfield{author}{\bibinfo{person}{FDA}.} \bibinfo{year}{2020}\natexlab{b}.
\newblock \bibinfo{title}{{Food \& Drug Administration. Digital Health
  Innovation Action Plan}}.
\newblock
\newblock
\newblock
\shownote{Available at https://www.fda.gov/medical-devices/digital-health.}


\bibitem[\protect\citeauthoryear{Grant, Moshyk, Diab, Caron, de~Lorenzi,
  Bisson, Menard, Lefebvre, Gauthier, Grondin, et~al\mbox{.}}{Grant
  et~al\mbox{.}}{2006}]%
        {grant2006integrating}
\bibfield{author}{\bibinfo{person}{Andrew Grant}, \bibinfo{person}{Andriy
  Moshyk}, \bibinfo{person}{Hassan Diab}, \bibinfo{person}{Philippe Caron},
  \bibinfo{person}{Fabien de Lorenzi}, \bibinfo{person}{Guy Bisson},
  \bibinfo{person}{Line Menard}, \bibinfo{person}{Richard Lefebvre},
  \bibinfo{person}{Patricia Gauthier}, \bibinfo{person}{Richard Grondin},
  {et~al\mbox{.}}} \bibinfo{year}{2006}\natexlab{}.
\newblock \showarticletitle{Integrating feedback from a clinical data warehouse
  into practice organisation}.
\newblock \bibinfo{journal}{\emph{International journal of medical
  informatics}} \bibinfo{volume}{75}, \bibinfo{number}{3-4}
  (\bibinfo{year}{2006}), \bibinfo{pages}{232--239}.
\newblock


\bibitem[\protect\citeauthoryear{Han, Haihong, Le, and Du}{Han
  et~al\mbox{.}}{2011}]%
        {han2011survey}
\bibfield{author}{\bibinfo{person}{Jing Han}, \bibinfo{person}{E Haihong},
  \bibinfo{person}{Guan Le}, {and} \bibinfo{person}{Jian Du}.}
  \bibinfo{year}{2011}\natexlab{}.
\newblock \showarticletitle{Survey on NoSQL database}. In
  \bibinfo{booktitle}{\emph{2011 6th international conference on pervasive
  computing and applications}}. IEEE, \bibinfo{pages}{363--366}.
\newblock


\bibitem[\protect\citeauthoryear{He, Zhang, Ren, and Sun}{He
  et~al\mbox{.}}{2016}]%
        {he2016deep}
\bibfield{author}{\bibinfo{person}{Kaiming He}, \bibinfo{person}{Xiangyu
  Zhang}, \bibinfo{person}{Shaoqing Ren}, {and} \bibinfo{person}{Jian Sun}.}
  \bibinfo{year}{2016}\natexlab{}.
\newblock \showarticletitle{Deep residual learning for image recognition}. In
  \bibinfo{booktitle}{\emph{Proceedings of the IEEE conference on computer
  vision and pattern recognition}}. \bibinfo{pages}{770--778}.
\newblock


\bibitem[\protect\citeauthoryear{Huang, Mankovich, Taira, Cho, Stewart, Ho,
  Chan, and Ishimitsu}{Huang et~al\mbox{.}}{1988}]%
        {huang1988picture}
\bibfield{author}{\bibinfo{person}{HK Huang}, \bibinfo{person}{NJ Mankovich},
  \bibinfo{person}{RK Taira}, \bibinfo{person}{PS Cho}, \bibinfo{person}{BK
  Stewart}, \bibinfo{person}{BK Ho}, \bibinfo{person}{KK Chan}, {and}
  \bibinfo{person}{Y Ishimitsu}.} \bibinfo{year}{1988}\natexlab{}.
\newblock \showarticletitle{{Picture archiving and communication systems (PACS)
  for radiological images: state of the art.}}
\newblock \bibinfo{journal}{\emph{Critical reviews in diagnostic imaging}}
  \bibinfo{volume}{28}, \bibinfo{number}{4} (\bibinfo{year}{1988}),
  \bibinfo{pages}{383--427}.
\newblock


\bibitem[\protect\citeauthoryear{Ketkar}{Ketkar}{2017}]%
        {ketkar2017introduction}
\bibfield{author}{\bibinfo{person}{Nikhil Ketkar}.}
  \bibinfo{year}{2017}\natexlab{}.
\newblock \showarticletitle{Introduction to pytorch}.
\newblock In \bibinfo{booktitle}{\emph{Deep learning with python}}.
  \bibinfo{publisher}{Springer}, \bibinfo{pages}{195--208}.
\newblock


\bibitem[\protect\citeauthoryear{Kim, Jang, Kim, Shin, and Park}{Kim
  et~al\mbox{.}}{2019}]%
        {kim2019design}
\bibfield{author}{\bibinfo{person}{Dong~Wook Kim}, \bibinfo{person}{Hye~Young
  Jang}, \bibinfo{person}{Kyung~Won Kim}, \bibinfo{person}{Youngbin Shin},
  {and} \bibinfo{person}{Seong~Ho Park}.} \bibinfo{year}{2019}\natexlab{}.
\newblock \showarticletitle{Design characteristics of studies reporting the
  performance of artificial intelligence algorithms for diagnostic analysis of
  medical images: results from recently published papers}.
\newblock \bibinfo{journal}{\emph{Korean journal of radiology}}
  \bibinfo{volume}{20}, \bibinfo{number}{3} (\bibinfo{year}{2019}),
  \bibinfo{pages}{405--410}.
\newblock


\bibitem[\protect\citeauthoryear{Lin, Goyal, Girshick, He, and Doll{\'a}r}{Lin
  et~al\mbox{.}}{2017}]%
        {lin2017focal}
\bibfield{author}{\bibinfo{person}{Tsung-Yi Lin}, \bibinfo{person}{Priya
  Goyal}, \bibinfo{person}{Ross Girshick}, \bibinfo{person}{Kaiming He}, {and}
  \bibinfo{person}{Piotr Doll{\'a}r}.} \bibinfo{year}{2017}\natexlab{}.
\newblock \showarticletitle{Focal loss for dense object detection}. In
  \bibinfo{booktitle}{\emph{Proceedings of the IEEE international conference on
  computer vision}}. \bibinfo{pages}{2980--2988}.
\newblock


\bibitem[\protect\citeauthoryear{Lin, Maire, Belongie, Hays, Perona, Ramanan,
  Doll{\'a}r, and Zitnick}{Lin et~al\mbox{.}}{2014}]%
        {lin2014microsoft}
\bibfield{author}{\bibinfo{person}{Tsung-Yi Lin}, \bibinfo{person}{Michael
  Maire}, \bibinfo{person}{Serge Belongie}, \bibinfo{person}{James Hays},
  \bibinfo{person}{Pietro Perona}, \bibinfo{person}{Deva Ramanan},
  \bibinfo{person}{Piotr Doll{\'a}r}, {and} \bibinfo{person}{C~Lawrence
  Zitnick}.} \bibinfo{year}{2014}\natexlab{}.
\newblock \showarticletitle{Microsoft coco: Common objects in context}. In
  \bibinfo{booktitle}{\emph{European conference on computer vision}}. Springer,
  \bibinfo{pages}{740--755}.
\newblock


\bibitem[\protect\citeauthoryear{Merkel}{Merkel}{2014}]%
        {merkel2014docker}
\bibfield{author}{\bibinfo{person}{Dirk Merkel}.}
  \bibinfo{year}{2014}\natexlab{}.
\newblock \showarticletitle{Docker: lightweight linux containers for consistent
  development and deployment}.
\newblock \bibinfo{journal}{\emph{Linux journal}} \bibinfo{volume}{2014},
  \bibinfo{number}{239} (\bibinfo{year}{2014}), \bibinfo{pages}{2}.
\newblock


\bibitem[\protect\citeauthoryear{Militello and Hutton}{Militello and
  Hutton}{1998}]%
        {militello1998applied}
\bibfield{author}{\bibinfo{person}{Laura~G Militello} {and}
  \bibinfo{person}{Robert~JB Hutton}.} \bibinfo{year}{1998}\natexlab{}.
\newblock \showarticletitle{{Applied cognitive task analysis (ACTA): a
  practitioner's toolkit for understanding cognitive task demands}}.
\newblock \bibinfo{journal}{\emph{Ergonomics}} \bibinfo{volume}{41},
  \bibinfo{number}{11} (\bibinfo{year}{1998}), \bibinfo{pages}{1618--1641}.
\newblock


\bibitem[\protect\citeauthoryear{Mongkolwat, Kleper, Talbot, and
  Rubin}{Mongkolwat et~al\mbox{.}}{2014}]%
        {mongkolwat2014national}
\bibfield{author}{\bibinfo{person}{Pattanasak Mongkolwat},
  \bibinfo{person}{Vladimir Kleper}, \bibinfo{person}{Skip Talbot}, {and}
  \bibinfo{person}{Daniel Rubin}.} \bibinfo{year}{2014}\natexlab{}.
\newblock \showarticletitle{The National Cancer Informatics Program (NCIP)
  Annotation and Image Markup (AIM) Foundation model}.
\newblock \bibinfo{journal}{\emph{Journal of digital imaging}}
  \bibinfo{volume}{27}, \bibinfo{number}{6} (\bibinfo{year}{2014}),
  \bibinfo{pages}{692--701}.
\newblock


\bibitem[\protect\citeauthoryear{Noorbakhsh-Sabet, Zand, Zhang, and
  Abedi}{Noorbakhsh-Sabet et~al\mbox{.}}{2019}]%
        {noorbakhsh2019artificial}
\bibfield{author}{\bibinfo{person}{Nariman Noorbakhsh-Sabet},
  \bibinfo{person}{Ramin Zand}, \bibinfo{person}{Yanfei Zhang}, {and}
  \bibinfo{person}{Vida Abedi}.} \bibinfo{year}{2019}\natexlab{}.
\newblock \showarticletitle{Artificial intelligence transforms the future of
  healthcare}.
\newblock \bibinfo{journal}{\emph{The American journal of medicine}}
  (\bibinfo{year}{2019}).
\newblock


\bibitem[\protect\citeauthoryear{Pantanowitz, Sharma, Carter, Kurc, Sussman,
  and Saltz}{Pantanowitz et~al\mbox{.}}{2018}]%
        {pantanowitz2018twenty}
\bibfield{author}{\bibinfo{person}{Liron Pantanowitz}, \bibinfo{person}{Ashish
  Sharma}, \bibinfo{person}{Alexis~B Carter}, \bibinfo{person}{Tahsin Kurc},
  \bibinfo{person}{Alan Sussman}, {and} \bibinfo{person}{Joel Saltz}.}
  \bibinfo{year}{2018}\natexlab{}.
\newblock \showarticletitle{Twenty years of digital pathology: An overview of
  the road travelled, what is on the horizon, and the emergence of
  vendor-neutral archives}.
\newblock \bibinfo{journal}{\emph{Journal of pathology informatics}}
  \bibinfo{volume}{9} (\bibinfo{year}{2018}).
\newblock


\bibitem[\protect\citeauthoryear{Parisot}{Parisot}{1995}]%
        {parisot1995dicom}
\bibfield{author}{\bibinfo{person}{Charles Parisot}.}
  \bibinfo{year}{1995}\natexlab{}.
\newblock \showarticletitle{The DICOM standard}.
\newblock \bibinfo{journal}{\emph{The International Journal of Cardiac
  Imaging}} \bibinfo{volume}{11}, \bibinfo{number}{3} (\bibinfo{year}{1995}),
  \bibinfo{pages}{171--177}.
\newblock


\bibitem[\protect\citeauthoryear{Pianykh}{Pianykh}{2009}]%
        {pianykh2009digital}
\bibfield{author}{\bibinfo{person}{Oleg~S Pianykh}.}
  \bibinfo{year}{2009}\natexlab{}.
\newblock \bibinfo{booktitle}{\emph{{Digital imaging and communications in
  medicine (DICOM): a practical introduction and survival guide}}}.
\newblock \bibinfo{publisher}{Springer Science \& Business Media}.
\newblock


\bibitem[\protect\citeauthoryear{Soremekun, Takayesu, and Bohan}{Soremekun
  et~al\mbox{.}}{2011}]%
        {soremekun2011framework}
\bibfield{author}{\bibinfo{person}{Olanrewaju~A Soremekun},
  \bibinfo{person}{James~K Takayesu}, {and} \bibinfo{person}{Stephen~J Bohan}.}
  \bibinfo{year}{2011}\natexlab{}.
\newblock \showarticletitle{Framework for analyzing wait times and other
  factors that impact patient satisfaction in the emergency department}.
\newblock \bibinfo{journal}{\emph{The Journal of emergency medicine}}
  \bibinfo{volume}{41}, \bibinfo{number}{6} (\bibinfo{year}{2011}),
  \bibinfo{pages}{686--692}.
\newblock


\bibitem[\protect\citeauthoryear{Subbaswamy and Saria}{Subbaswamy and
  Saria}{2020}]%
        {subbaswamy2020development}
\bibfield{author}{\bibinfo{person}{Adarsh Subbaswamy} {and}
  \bibinfo{person}{Suchi Saria}.} \bibinfo{year}{2020}\natexlab{}.
\newblock \showarticletitle{From development to deployment: dataset shift,
  causality, and shift-stable models in health AI}.
\newblock \bibinfo{journal}{\emph{Biostatistics}} \bibinfo{volume}{21},
  \bibinfo{number}{2} (\bibinfo{year}{2020}), \bibinfo{pages}{345--352}.
\newblock


\bibitem[\protect\citeauthoryear{Taylor, Mielke, and Mongan}{Taylor
  et~al\mbox{.}}{2018}]%
        {taylor2018automated}
\bibfield{author}{\bibinfo{person}{Andrew~G Taylor}, \bibinfo{person}{Clinton
  Mielke}, {and} \bibinfo{person}{John Mongan}.}
  \bibinfo{year}{2018}\natexlab{}.
\newblock \showarticletitle{Automated detection of moderate and large
  pneumothorax on frontal chest X-rays using deep convolutional neural
  networks: A retrospective study}.
\newblock \bibinfo{journal}{\emph{PLoS medicine}} \bibinfo{volume}{15},
  \bibinfo{number}{11} (\bibinfo{year}{2018}).
\newblock


\bibitem[\protect\citeauthoryear{Trinks and Felden}{Trinks and Felden}{2017}]%
        {trinks2017real}
\bibfield{author}{\bibinfo{person}{Sebastian Trinks} {and}
  \bibinfo{person}{Carsten Felden}.} \bibinfo{year}{2017}\natexlab{}.
\newblock \showarticletitle{Real time analytics—State of the art: Potentials
  and limitations in the smart factory}. In \bibinfo{booktitle}{\emph{2017 IEEE
  International Conference on Big Data (Big Data)}}. IEEE,
  \bibinfo{pages}{4843--4845}.
\newblock


\bibitem[\protect\citeauthoryear{Venkatesh, Thong, and Xu}{Venkatesh
  et~al\mbox{.}}{2016}]%
        {venkatesh2016unified}
\bibfield{author}{\bibinfo{person}{Viswanath Venkatesh},
  \bibinfo{person}{James~YL Thong}, {and} \bibinfo{person}{Xin Xu}.}
  \bibinfo{year}{2016}\natexlab{}.
\newblock \showarticletitle{Unified theory of acceptance and use of technology:
  A synthesis and the road ahead}.
\newblock \bibinfo{journal}{\emph{Journal of the Association for Information
  Systems}} \bibinfo{volume}{17}, \bibinfo{number}{5} (\bibinfo{year}{2016}),
  \bibinfo{pages}{328--376}.
\newblock


\bibitem[\protect\citeauthoryear{Warnock, Toland, Evans, Wallace, and
  Nagy}{Warnock et~al\mbox{.}}{2007}]%
        {warnock2007benefits}
\bibfield{author}{\bibinfo{person}{Max~J Warnock}, \bibinfo{person}{Christopher
  Toland}, \bibinfo{person}{Damien Evans}, \bibinfo{person}{Bill Wallace},
  {and} \bibinfo{person}{Paul Nagy}.} \bibinfo{year}{2007}\natexlab{}.
\newblock \showarticletitle{Benefits of using the DCM4CHE DICOM archive}.
\newblock \bibinfo{journal}{\emph{Journal of Digital Imaging}}
  \bibinfo{volume}{20}, \bibinfo{number}{1} (\bibinfo{year}{2007}),
  \bibinfo{pages}{125--129}.
\newblock


\bibitem[\protect\citeauthoryear{Zech, Badgeley, Liu, Costa, Titano, and
  Oermann}{Zech et~al\mbox{.}}{2018}]%
        {zech2018variable}
\bibfield{author}{\bibinfo{person}{John~R Zech}, \bibinfo{person}{Marcus~A
  Badgeley}, \bibinfo{person}{Manway Liu}, \bibinfo{person}{Anthony~B Costa},
  \bibinfo{person}{Joseph~J Titano}, {and} \bibinfo{person}{Eric~Karl
  Oermann}.} \bibinfo{year}{2018}\natexlab{}.
\newblock \showarticletitle{Variable generalization performance of a deep
  learning model to detect pneumonia in chest radiographs: a cross-sectional
  study}.
\newblock \bibinfo{journal}{\emph{PLoS medicine}} \bibinfo{volume}{15},
  \bibinfo{number}{11} (\bibinfo{year}{2018}).
\newblock


\end{thebibliography}

\end{document}